\documentclass[prl,letterpaper,twocolumn,showpacs,superscriptaddress,amsmath,amsfonts,amssymb,preprintnumbers]{revtex4}
\pdfoutput=1
\usepackage[T1]{fontenc}\usepackage[latin1]{inputenc}
\usepackage{dcolumn,graphicx,color,booktabs,microtype,afterpage} 
\usepackage[charter]{mathdesign}

\renewcommand{\tablename}{Table}
\makeatletter\renewcommand{\fnum@table}[1]{\tablename~\thetable.}\makeatother

\usepackage[colorlinks,plainpages=false,linkcolor=blue,urlcolor=blue,citecolor=black,pdfpagemode=UseNone,pdfstartview=FitBH]{hyperref}

\newcommand{ \ybco }{\mbox{YBa$_2$Cu$_3$O$_{6+y}$}}

\newcommand{ \ybcosixfour }{\mbox{YBa$_2$Cu$_3$O$_{6.45}$}}
\newcommand{ \ybcosixfive }{\mbox{YBa$_2$Cu$_3$O$_{6.5}$}}
\newcommand{ \ybcosixsix }{\mbox{YBa$_2$Cu$_3$O$_{6.6}$}}

\newcommand{ \ybcoonetwofour }{\mbox{YBa$_2$Cu$_4$O$_{8}$}}

\newcommand{ \lsco }{\mbox{La$_{2-x}$Sr$_x$CuO$_4$}}

\newcommand{ \ARPES }{\mbox{ARPES}}

\newcommand{ \cuot }{\mbox{CuO$_2$}}

\newcommand{ \ph }{\mbox{$p_{\text h}$}}

\newcommand{ \ie }{\mbox{i.\,e.}}

\usepackage{varioref}
\usepackage{amssymb}

\begin{document}


\title{Magnetic-Field-Enhanced Incommensurate Magnetic Order in the Underdoped High-Temperature Superconductor YBa$_2$Cu$_3$O$_{6.45}$}

\author{D. Haug}
\affiliation{Max-Planck-Institut f{\"u}r Festk{\"o}rperforschung,
Heisenbergstra{\ss}e 1, D-70569 Stuttgart, Germany}

\author{V. Hinkov}
\email[email:]{V.Hinkov@fkf.mpg.de} \affiliation{Max-Planck-Institut
f{\"u}r Festk{\"o}rperforschung, Heisenbergstra{\ss}e 1, D-70569
Stuttgart, Germany}

\author{A. Suchaneck}
\affiliation{Max-Planck-Institut f{\"u}r Festk{\"o}rperforschung,
Heisenbergstra{\ss}e 1, D-70569 Stuttgart, Germany}

\author{D. S. Inosov}
\affiliation{Max-Planck-Institut f{\"u}r Festk{\"o}rperforschung,
Heisenbergstra{\ss}e 1, D-70569 Stuttgart, Germany}

\author{N. B. Christensen}
\affiliation{Laboratory for Neutron Scattering, ETH Z{\"u}rich \&
Paul Scherrer Institut, CH-5232 Villigen, Switzerland}
\affiliation{Materials Research Division, Ris{\o} DTU, Technical
University of Denmark, DK-4000 Roskilde, Denmark}
\affiliation{Nano-Science Center, Niels Bohr Institute, University
of Copenhagen, DK-2100 Copenhagen, Denmark}

\author{Ch. Niedermayer}
\affiliation{Laboratory for Neutron Scattering, ETH Z{\"u}rich \&
Paul Scherrer Institut, CH-5232 Villigen, Switzerland}

\author{P. Bourges}
\affiliation{Laboratoire L{\'e}on Brillouin, CEA-CNRS, CEA-Saclay,
F-91191 Gif-s{\^u}r-Yvette, France}

\author{Y. Sidis}
\affiliation{Laboratoire L{\'e}on Brillouin, CEA-CNRS, CEA-Saclay,
F-91191 Gif-s{\^u}r-Yvette, France}

\author{J. T. Park}
\affiliation{Max-Planck-Institut f{\"u}r Festk{\"o}rperforschung,
Heisenbergstra{\ss}e 1, D-70569 Stuttgart, Germany}

\author{A. Ivanov}
\affiliation{Institut Laue-Langevin, 6 Rue Jules Horowitz, F-38042
Grenoble Cedex 9, France}

\author{C. T. Lin}
\affiliation{Max-Planck-Institut f{\"u}r Festk{\"o}rperforschung,
Heisenbergstra{\ss}e 1, D-70569 Stuttgart, Germany}

\author{J. Mesot}
\affiliation{Laboratory for Neutron Scattering, ETH Z{\"u}rich \&
Paul Scherrer Institut, CH-5232 Villigen, Switzerland}
\affiliation{Institut de Physique de la Mati{\`e}re Complexe, EPFL,
CH-1015 Lausanne,  Switzerland}

\author{B. Keimer}
\affiliation{Max-Planck-Institut f{\"u}r Festk{\"o}rperforschung,
Heisenbergstra{\ss}e 1, D-70569 Stuttgart, Germany}

\date{\today}

\begin{abstract}
We present a neutron-scattering study of the static and dynamic spin
correlations in the underdoped high-temperature superconductor YBa$_2$Cu$_3$O$_{6.45}$ in magnetic fields up to 15 T. The field strongly enhances static incommensurate magnetic order at low temperatures and induces a spectral-weight shift in the magnetic-excitation spectrum. A reconstruction of the Fermi surface driven by the field-enhanced magnetic superstructure may thus be responsible for the unusual Fermi surface topology revealed by recent quantum-oscillation experiments.
\end{abstract}

\pacs{\vspace{-0.2em}74.25.Ha 74.72.Bk 78.70.Nx 71.18.+y \vspace{-0.5em}}


\maketitle

The Landau theory, which treats correlated electron systems as a
liquid of weakly interacting quasiparticles, is one of the central
tenets underlying our current microscopic understanding of metals.
For a long time, the existence of well-defined Landau quasiparticles
in underdoped high-temperature superconductors had been called into
question, based in part on the unusual thermodynamic and transport
properties of these materials. Recently, however, the discovery of quantum oscillations in \ybcosixfive\ (Refs. \onlinecite{Doiron07,LeBoeuf07,Jaudet08}) and its close cousin \ybcoonetwofour\ (Ref. \onlinecite{Yelland08}), the two underdoped cuprates least affected by disorder due to randomly placed dopant atoms, has confirmed the presence of coherent fermionic quasiparticles.

While resolving a major open question in research on
high-temperature superconductivity, the quantum-oscillation
experiments have generated a new set of puzzles. The observed
oscillation frequencies, combined with the negative sign of the
low-temperature Hall coefficient, are indicative of small
electron-like Fermi surface pockets. A straightforward electron
count then shows that a complex Fermi surface comprising both
electron and hole pockets is required to satisfy Luttinger's
theorem. Very recent quantum-oscillation data have indeed provided
evidence of the hole pockets inferred from these arguments
\cite{Sebastian08}. However, neither angle-resolved photo\-emission
spectroscopy (\ARPES) data on \ybco\ \cite{Hossain08,Zabolotnyy07}
(or any other hole-doped high-temperature superconductor) nor
ab-initio band structure calculations for \ybcosixfive\ and
\ybcoonetwofour\ (Refs. \onlinecite{Carrington07,Elfimov08}) show clear signs of a
Fermi surface topology comprised of small electron and hole pockets.

Keeping in mind that quantum-oscillation experiments require
magnetic fields $H \sim 50$ T, above the upper critical field for
superconductivity, and that \ARPES\ measurements can only be
performed for $H=0$, a Fermi surface reconstruction due to a
magnetic-field-induced superstructure offers a possible solution to
this puzzle. Indeed, field-induced commensurate \cite{Chen08} and
incommensurate \cite{Harrison09} spin modulations as well as striped
\cite{Millis07} and $d$-density-wave \cite{Chakravarty08} states
have been invoked as explanations of the small Fermi surface
pockets. We use elastic and inelastic neutron scattering to
demonstrate that \ybco\ crystals with $y=0.45$ (corresponding to
doping levels close to those studied in the quantum oscillation
experiments) develop robust incommensurate magnetic order in high
magnetic fields, thus providing strong experimental support for one
of the proposed scenarios as well as a perspective for a unified
description of the spin correlations and fermiology in the
underdoped cuprates.

The neutron scattering experiments were performed on the same array
of untwinned \ybcosixfour\ single crystals ($T_c=35$ K) recently
shown to exhibit liquid-crystal-like \cite{Kivelson98}
incommensurate spin fluctuations (with a propagation direction
selected by the small orthorhombic distortion of the crystal
structure) below $\sim 150$ K for $H=0$, while muon-spin-rotation
($\mu$SR) experiments show static magnetic order only below
\mbox{$\sim 2$ K} (Ref. \onlinecite{Hinkov08}). The array consists
of $\sim 100$ almost twin-free \ybcosixfour\ single crystals with a
total mass of $\sim 2.0$ g and a majority twin domain population of
$>$ 92\%. The crystals were individually characterized by
magnetometry and have a transition temperature of $T_c=35$ K with a
transition width of 2--3~K. The lattice parameters $a=3.8388
\text{\,\AA}$, $b=3.8747 \text{\,\AA}$, and $c=11.761 \text{\,\AA}$
correspond to a hole doping level \ph\ = 0.085 $\pm$ 0.01 per Cu ion
\cite{Liang06}. Further details about the sample and the detwinning
procedure are given in Refs. \onlinecite{Hinkov08} and
\onlinecite{Hinkov07}.


The experiments were carried out at the triple-axis spectrometer IN14 at the Institut Laue-Langevin. Pyrolytic graphite crystals, set for the (002) reflection, were used to monochromate and analyze the neutron beam. The final
wave vector was fixed to $k_\textup{f}=1.5 \text{\,\AA}^{-1}$. No
collimation was used to maximize the neutron flux, and a beryllium
filter extinguished higher-order contaminations of the neutron beam.

The sample was mounted in a 15-Tesla vertical-field cryomagnet.
The scattering plane was spanned by the vectors (1 0 0) and (0 1 2). Throughout this paper, the wave vector $\mathbf{Q}=(Q_\mathrm{H},Q_\mathrm{K},Q_\mathrm{L})$ is quoted in r.l.u., \ie\ in units of the reciprocal lattice vectors $\mathbf{a^*}$, $\mathbf{b^*}$ and $\mathbf{c^*}$ ($a^*=2\pi/a$ etc.).
The scattering geometry implies an angle of $\sim33^{\circ}$ between
$\mathbf{H}$ and $\bf c$. For the field-dependent experiments, the external magnetic field was thus applied
mainly perpendicular to the \cuot\ planes. All scans for $H \neq 0$ were performed
after field cooling.

\begin{figure}[]
\includegraphics[width=0.48\textwidth]{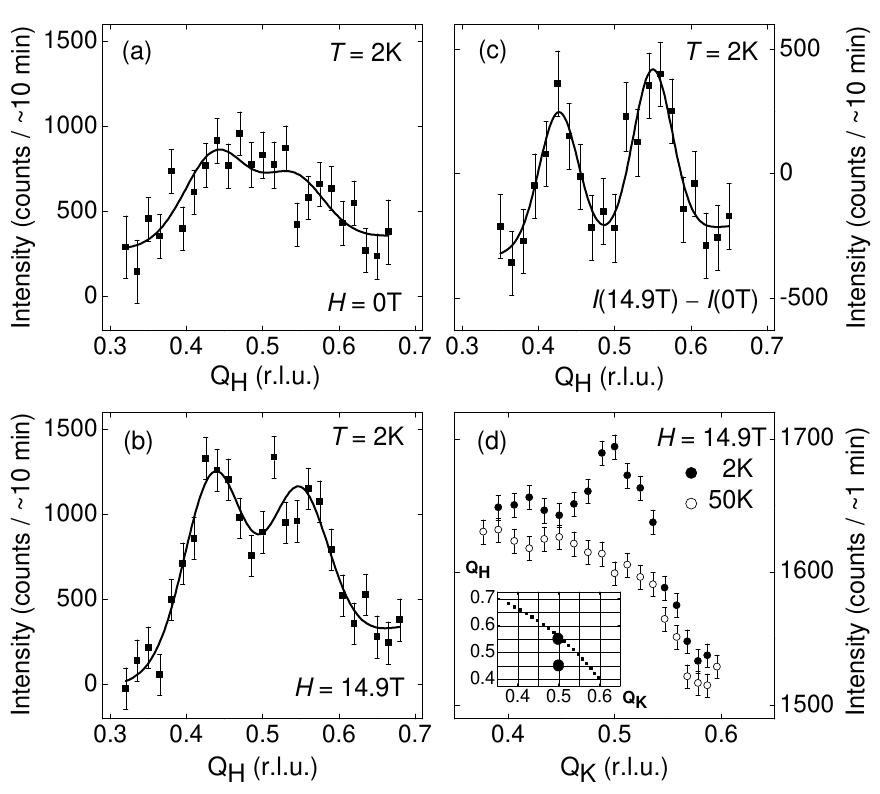}
\caption{\label{figQscans} (a)--(c) Elastic scans along
$Q_\mathrm{H}$ at fixed $Q_\mathrm{K} = 0.5$ and $Q_\mathrm{L} = 1$.
The lines are results of two-Gaussian fits to the data. Panels (a)
and (b) compare the incommensurate spin modulation at $H=0$ and at
$H=14.9$~T at low temperatures ($T=2$~K). The background at $T=50$~K
was subtracted. Panel (c) visualizes the effect of applying
$H=14.9$~T at $T=2$~K by subtracting the intensity at $H=0$ from the
intensity at $H=14.9$~T. (d) Raw (uncorrected) rocking scan profile
at $H=14.9$~T through the incommensurate peak at (0.56, 0.5, 1)
along the reciprocal-space trajectory indicated in the inset. Closed
circles indicate the intensity at 2~K, open circles the intensity at
50~K.
}
\end{figure}

We first focus on the magnetic response obtained by setting the
neutron spectrometer for energy transfer $E = 0$. In agreement with
previous data on \ybcosixfour\ (Ref. \onlinecite{Hinkov08}), scans taken at temperatures below $\sim 30$ K for $H=0$ in this configuration and shown
in Fig. \ref{figQscans}(a) exhibit a small incommensurate signal with amplitude
$\sim 0.05$~$\mu_B$ per planar Cu site at $T=2$ K, estimated from
$\mu$SR measurements on the same sample \cite{Hinkov08}. The finite
energy resolution $\Delta E\sim 0.2$ meV of the instrument implies
that static order as well as slow fluctuations (with energies less
than $\Delta E$) contribute to this response. However, $\mu$SR
experiments show that at temperatures around 2 K
the characteristic energy scale for fluctuations of the electron
spin system is in the $\mu$eV range. Since this is well below the
characteristic energy scale \mbox{$\hbar \omega_c \sim 1$ meV} for
cyclotron motion of the electrons at $H \sim 50$ T
\cite{Doiron07,Jaudet08}, the $E=0$ data can be regarded as a
manifestation of static incommensurate magnetic order when
discussing implications for the quantum-oscillation data.

\begin{figure}
\includegraphics[width=0.42\textwidth]{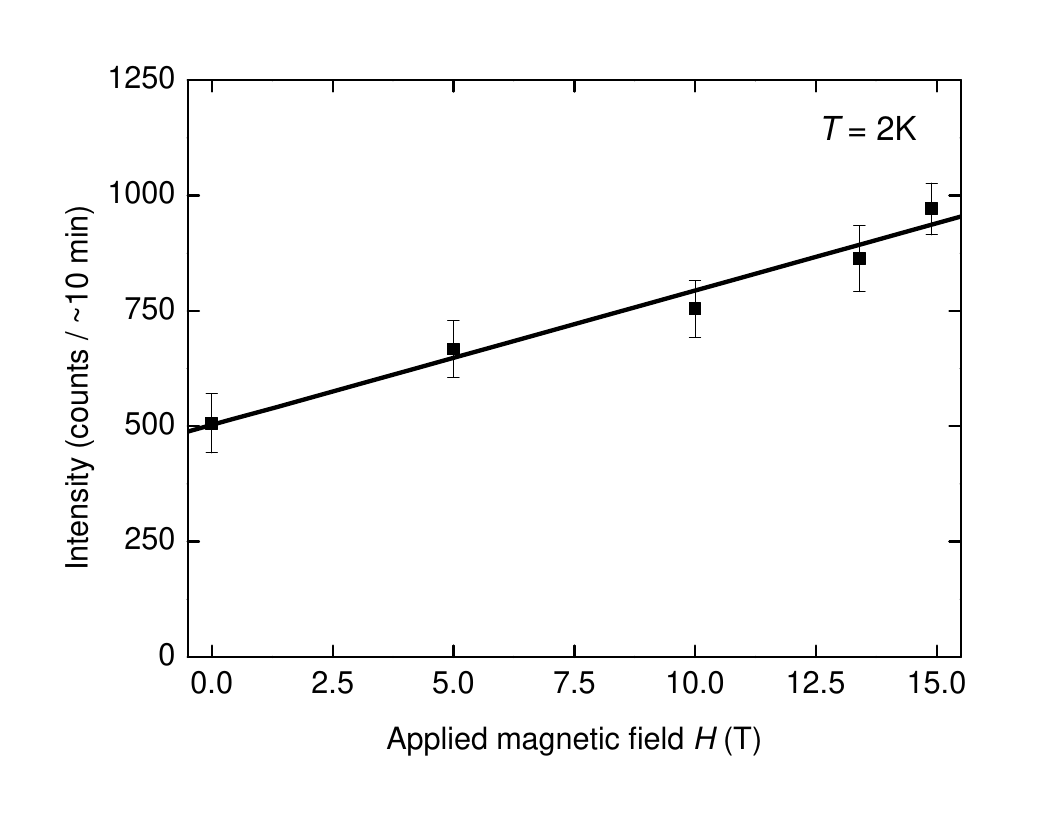}
\caption{\label{figFieldDep} Dependence of the elastic peak
intensity at the incommensurate position (0.56, 0.5, 1) on
the external field at $T=2$~K. The data was extracted from the scans
shown in Fig. \ref{figQscans} and scans similar to those. The line
is the result of a linear fit to the data points.}
\end{figure}

Corresponding elastic data at the highest available field, $H=14.9$
T, demonstrate that the ordered moment increases strongly with
magnetic field, Fig. \ref{figQscans}(b),(c). In order to obtain a
quantitative description of the evolution of the incommensurate spin
modulation with $H$, the data were fitted to Gaussian profiles
(lines in Fig.  \ref{figQscans}). Neither the resolution-corrected
peak width ($0.09\pm 0.01$~r.l.u. full width at half maximum) nor
the incommensurability ($0.049\pm 0.003$ at $H=0$, $0.055\pm
0.003$~r.l.u. at $H=14.9$ T) vary significantly with $H$. However, the integrated intensity of the elastic peak extracted from these fits increases continuously with $H$, Fig. \ref{figFieldDep}. At $H=14.9$ T, it is about a factor of two larger than the zero-field intensity.

\begin{figure}[h]
\includegraphics[width=0.3\textwidth]{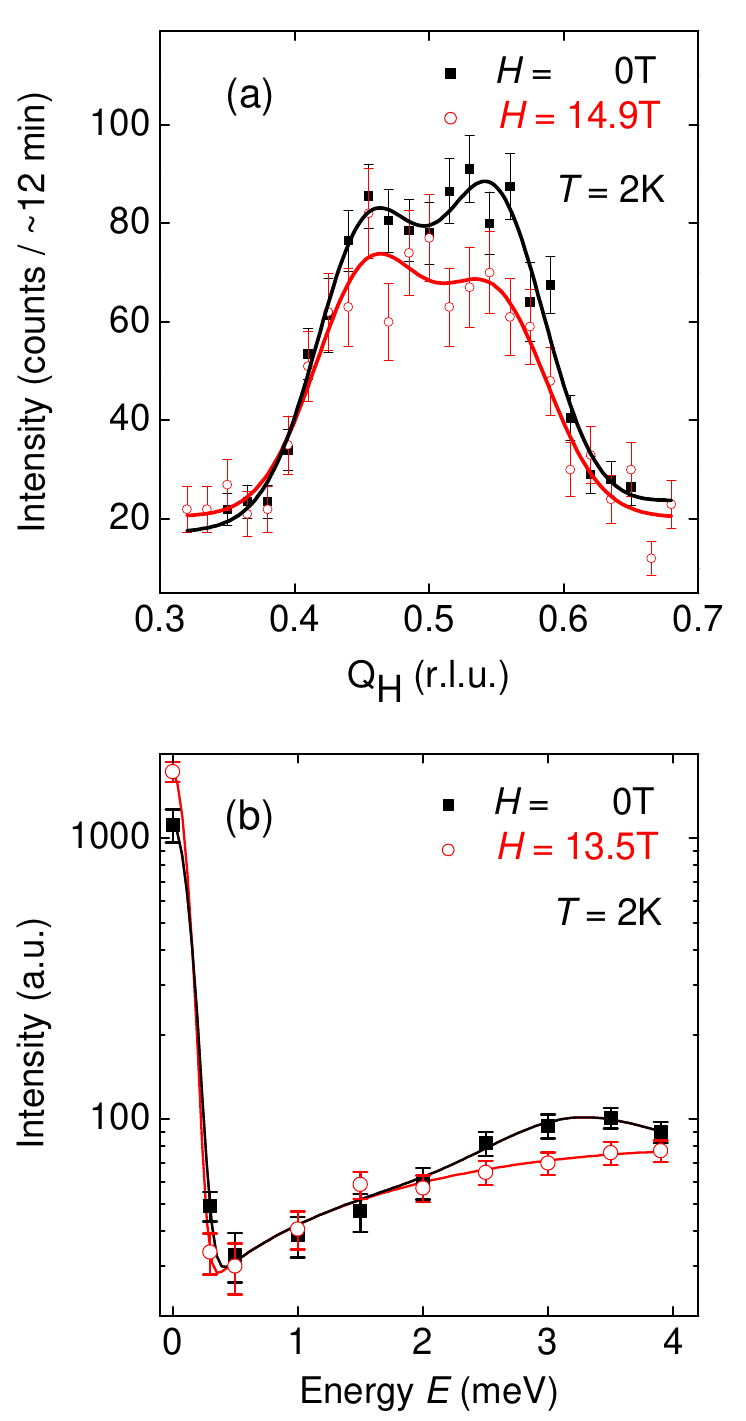}
\caption{\label{figInelastic} (color online). (a) Inelastic scans at
$E=3$~meV and $T=2$~K along $Q_\mathrm{H}$ at fixed
$Q_\mathrm{K}=0.5$ and $Q_\mathrm{L}=1$. Black squares indicate the
raw scattering intensity at $H=0$, open circles the intensity at
$H=14.9$ T. The lines represent the results of two-Gaussian fits to
the data. (b) Energy dependence of the magnetic peak
intensity at the incommensurate position (0.56, 0.5, 1) in zero
field (closed squares) and in $H=13.5$ T (open circles), corrected
for the background. Note the logarithmic intensity scale.}
\end{figure}

The total moment sum rule for the scattering function
\cite{Lorenzana05} stipulates that the spectral weight integrated
over all energies and momenta is conserved. This implies that the
amplitude of the inelastic response has to decrease in order to
compensate for the spectral weight accumulated in the elastic peak.
We have therefore studied the effect of the external magnetic field
on the low-energy spin excitations. The results are summarized in
Fig. \ref{figInelastic}. For an excitation energy $E = 3$ meV, the
spin fluctuation intensity decreases appreciably at $H=14.9$ T,
while the shape of the scattering profile remains unaffected, Fig.
\ref{figInelastic}(a). Surprisingly, the suppression with field is
most pronounced around 3--4 meV, Fig. \ref{figInelastic}(b);
excitations at lower $E$ are much less affected by the external
field. The detailed quantitative verification of the sum rule is
generally complicated for technical reasons restricting the energy
range which can be studied at a given spectrometer \cite{note}, and in cuprates in
addition by the low magnetic intensity \cite{Fong96, Tranquada04, Chang07}.
Nevertheless, a comparison of the integrated intensities provides valuable insight:
Our data indicate that the spectral weight reduction in the energy range
covered by our experiment ($E<4$~meV, \cite{note}), accounts for
roughly half of the gain in elastic intensity with increasing $H$.
Fig. \ref{figInelastic}(b) indicates that at least part of the remaining
spectral weight originates from a range of a few meV immediately above
our upper limit of 4~meV, while we cannot exclude the possibility that some
intensity might come from even higher energies.

Previous experiments on the \lsco\ family of high-temperature
superconductors have provided ample evidence of field-induced or
field-enhanced incommensurate magnetic order
\cite{Katano00,Lake02,Khaykovich02,Khaykovich05,Chang07,Chang08, Chang09}. A
field enhancement of the magnetic order has also been reported for some
electron-doped cuprates \cite{Kang05}, but the magnetic correlations
in this family are commensurate (with a propagation vector identical
to the undoped antiferromagnet) at all doping levels. Our
observation now indicates that field amplification of incommensurate
magnetic order is a universal property of the hole-doped cuprates.

However, we also point out some material-specific aspects of the
magnetic phase diagram. First, we note that saturation behavior of
the elastic intensity at high field is observed in \lsco\ when the
amplitude of the ordered moment reaches $\sim 0.2 \mu_B$
\cite{Lake02}. As the ordered moment at the highest field covered by
our experiment, $\sim 0.07 \mu_B$, is substantially lower, the
absence of saturation effects is not surprising. The field-induced
amplification of the peak intensity at $E = 0$ also generally agrees
with theoretical work on incommensurate spin-density wave order
competing with $d$-wave superconductivity \cite{Demler01}. Evidence
for the theoretically predicted logarithmic corrections to the
leading linear $H$-dependence is, however, not apparent in the data
shown in Fig. \ref{figFieldDep}.

Second, most of the field-dependent experiments on \lsco\ were
performed at doping levels around 1/8 hole per Cu atom. The doping
level of our \ybcosixfour\  is significantly lower (0.085 holes per
planar Cu), and the incommensurability of the magnetic response is
about a factor of two smaller than that of \lsco\ at comparable
doping levels.

Unlike the findings at $E = 0$, the field-induced response of the
low-energy spin excitations in \ybcosixfour\ is quite different from
corresponding observations in \lsco\ -- in particular, our data bear
no indication of a spin gap closing with field (Fig.
\ref{figInelastic}). To a large extent, we attribute this to the
different hole doping: It is sufficiently high in the investigated
\lsco\ samples as to result in a pronounced superconducting spin gap
\cite{Lake01, Tranquada04, Chang07, Chang09}. In contrast, our
sample develops no such gap below $T_c$ in the first place
\cite{Hinkov08}, and exhibits a robust elastic peak already at $H =
0$. Such a phenomenology is established in the \lsco\ samples only
at elevated fields $H$ (e.g. in Ref. \onlinecite{Chang09}). The main
effect of further increasing $H$ is an increase of the elastic
intensity, in analogy to what we observe in \ybcosixfour.

While the spectral-weight reduction around 3--4~meV (Fig.
\ref{figInelastic}) is reminiscent of the field-induced suppression
of the intensity of the ``resonance mode'' in more highly doped
\ybcosixsix\ \cite{Dai00}, attributed to a destabilization of the
superconducting state through orbital depairing, the close proximity
of our sample to the antiferromagnetic insulating state casts doubts
on the applicability of such concepts at our low doping level. It is
thus interesting to compare our data to the field-induced response
of an undoped Heisenberg antiferromagnet with planar exchange
anisotropy \cite{Tranquada89}. In this system, a field applied
perpendicularly to the planes hardens magnon excitations polarized
out of the plane while leaving the gapless in-plane excitations
unaffected. Spin polarization analysis of the inelastic neutron
intensity in the presence of magnetic field is required to assess
the validity of this concept.

In summary, we have shown that an external magnetic field strongly
enhances static incommensurate magnetic order and induces a spectral
weight rearrangement in the magnetic excitation spectrum of
\ybcosixfour. Based on previous studies of \lsco\
\cite{Katano00,Lake02,Khaykovich02,Khaykovich05,Chang07,Chang08,Chang09}, it
is reasonable to expect a similar effect at the slightly larger
doping level of \ybcosixfive, where the quantum oscillation
experiments were performed. On general grounds, the field-enhanced
magnetic superstructure is expected to drive a reconstruction of the
Fermi surface from the large hole barrels predicted by band
structure calculations towards a more complex topology comprised of
small electron and hole pockets. In agreement with this argument,
the multi-sheet Fermi surface observed in electron-doped
high-temperature superconductors is now generally attributed to the
influence of commensurate antiferromagnetism
\cite{Park07,Harrison07}, which has been independently observed by
neutron diffraction \cite{Kang05}. Recent \ARPES\ experiments on Nd-
and Eu-substituted \lsco\ have also revealed Fermi surface
reconstructions possibly associated with stripe order
\cite{Chang08b,Zabolotnyy08}. Based on an extrapolation of the
intensity of the incommensurate elastic peak of \ybcosixfour\ to 50
T, where quantum oscillations are observed to set in, we estimate an
ordered moment of $\sim 0.13 \mu_B$, which is comparable to the
value observed in the fully established stripe-ordered state in
Nd-doped \lsco\ \cite{Chang08} and the typical ordered moments in
superconducting electron-doped cuprates \cite{Kang05}.

It is therefore likely that the incommensurate magnetic order we
have found strongly influences the Fermi surface geometry determined
by the high-field quantum-oscillation experiments. Theoretical
calculations of the Fermi surface geometry in the presence of stripe
order with commensurate in-plane wave vector 1/8, which were
inspired by experiments in \lsco, support this conclusion and
demonstrate that this mechanism can yield quantum oscillation
frequencies of the correct order of magnitude \cite{Millis07}. Our
experiments now show that a generalization of this work to
incommensurate magnetic order with substantially smaller wave
vectors is required to obtain a quantitative description of the
quantum-oscillation data. In closing, we note that our experiments
only probe spin (and not charge) degrees of freedom, and do not
discriminate between longitudinal (stripe)
\cite{Millis07} and transverse (spiral) \cite{Milstein08}
modulations of the spin system. It therefore appears worthwhile to
explore the possible influence of both types of order on the
physical properties of \ybco\ at high magnetic fields.

\begin{acknowledgments}
We thank O. Sushkov and C. Bernhard for stimulating discussions, A. Hiess and M. B{\"o}hm for
support during the experiment at IN14, S. Lacher, B. Baum and H.
Wendel for their help in sample preparation, C. Busch and H. Bender
for technical support, and R. Dinnebier for determining the lattice
parameters using X-ray powder diffraction.
\end{acknowledgments}


\begin{thebibliography}{35}
\bibitem{note} In our experiment, to be able to study both the elastic peak and the low-energy spin excitations, we had to use a cold triple-axis spectrometer.
This, in conjunction with the bulky 15~T cryomagnet, limited the
energy range for geometrical reasons.
\end{thebibliography}

\end{document}